\documentclass[aps,pra,showpacs,amsmath,amssymb,twocolumn,superscriptaddress]{revtex4-1}
\pdfoutput=1

\usepackage{graphicx}
\usepackage{xcolor}
\usepackage{color}
\usepackage{amsmath}
\usepackage{afterpage}
\usepackage{cancel}
\usepackage[utf8]{inputenc}

\begin{document}

\title{Cross-Kerr nonlinearity in optomechanical systems}

\author{Rapha\"el Khan}
\email[]{raphael.khan@aalto.fi}
\affiliation{Low Temperature Laboratory, Aalto University, P.O. Box 15100, FI-00076 AALTO, Finland}
\author{F. Massel}
\email[]{francesco.p.massel@jyu.fi}
\author{T.~T.~Heikkil\"a}
\email[]{tero.t.heikkila@jyu.fi}
\affiliation{Department of Physics and Nanoscience Center, University of Jyva\"akyl\"a,
P.O. Box 35 (YFL), FI-40014 University of Jyv\"askyl\"a, Finland}
\newcommand{\tmpnote}[1]%
   {\begingroup{\it (FIXME: #1)}\endgroup}
   \newcommand{\comment}[1]%
       {\marginpar{\tiny C: #1}}

\begin{abstract}
We consider the response of a nanomechanical resonator interacting with an electromagnetic cavity via a radiation pressure coupling and a cross-Kerr coupling. Using a mean field approach we solve the dynamics of the system, and show the different corrections coming from the radiation pressure and the cross-Kerr effect to the usually considered linearized dynamics.
\end{abstract}
\maketitle
\section{Introduction}
Cavity optomechanics offers a framework to study the coupling between an electromagnetic field  and the vibrations of a mechanical resonator. The interaction between these two systems is usually mediated by a radiation-pressure type coupling proportional -- through a coupling constant $g$-- to the number of photons $n_c$ in the cavity and the displacement of the mechanical resonator. The radiation pressure coupling offers the possibility of altering the resonant frequency  of the mechanical resonator and its damping. The latter can be used for  cooling \cite{PhysRevLett.101.197203,PhysRevLett.97.243905,PhysRevLett.99.093902} or amplification \cite{massel_microwave_2011}. Moreover, the nonlinearity of the interaction may allow for the observation of macroscopic quantum phenomena such as quantum superpositon of states \cite{PhysRevLett.91.130401,PhysRevLett.98.030405} or quantum squeezed states \cite{1367-2630-10-9-095010}. The requirements for observing these quantum phenomena are the necessity of being close to the ground state and being in the strong coupling regime \cite{NatureGrob,NatureT} where $g$ is larger than the cavity and the mechanical resonator decay rate. However, $g$ is usually weak and to bypass this constraint a strong drive to the cavity is applied at the cost of losing the nonlinear property of the interaction.

Our recent proposal  \cite{PhysRevLett.112.203603}, in which the cavity and the resonator are coupled to a Josephson junction, shows that the interaction between the cavity and  the resonator can be  enhanced via the non-linearity of the  Josephson effect. Quadratic and higher-order interactions in the displacement have been investigated also in different setups such as membrane in the middle geometries \cite{Bhattacharya:2007hh,Thompson:2008dx,Xuereb:2013ug}.

Analogously to the setup mentioned above, the  non-linearity of the Josephson effect leads to an additional nonlinear interaction, namely a cross-Kerr coupling $g_{ck}$ between the cavity and the resonator. The difference resides in the fact that in the Josephson junction setup the relative value of $g_{ck}$ and $g$ depends on the value of the gate charge to a superconducting island, whereas in \cite{Thompson:2008dx, Xuereb:2013ug}, it generally reflects the position of the resonator within the cavity. 

In the context of optomechanical systems, the cross-Kerr coupling between the resonator and the cavity induces a change to the refractive index of the cavity depending on the \textit{number} of phonons in the resonator, whereas the radiation pressure coupling gives rise to an analogous effect, but depending on the \textit{displacement} of the mechanical resonator. 

In this paper we solve the dynamics of the cavity and the mechanical resonator in the presence of the cross-Kerr and the radiation pressure coupling. We determine the effects of the cross-Kerr coupling on the red and blue sidebands within a mean field approach. In particular, we demonstrate that  the sideband peak is shifted due to  the cross-Kerr coupling. In addition, the cross-Kerr coupling induces a nonmonotonuous response of the effective mechanical damping as a function of  the number of photons pumped into the cavity.
\begin{figure}[h]
\includegraphics[width=0.8\columnwidth]{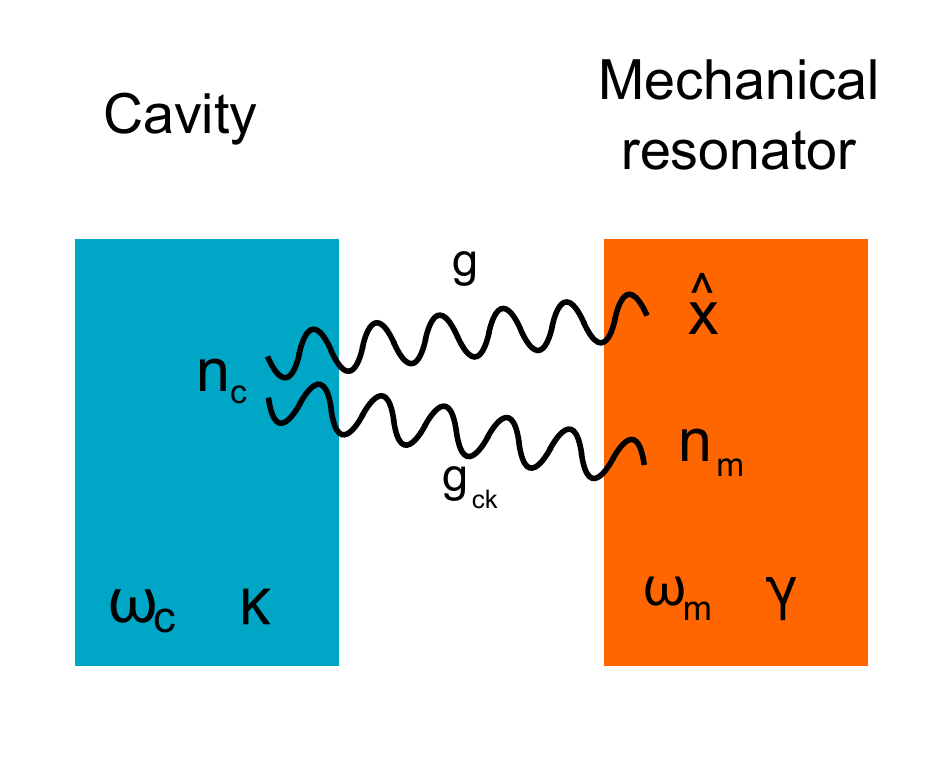}
\caption{Schematic picture of the system. A cavity and a mechanical resonator coupled via a radiation type coupling $g$ and a cross-Kerr coupling $g_{ck}$. The number of photons in the cavity $n_c$ is coupled to the oscillations of the mechanical resonator $\hat{x}$  and  the number of phonons in the mechanical resonator $n_m$. }
\label{Fig:schema}
\end{figure}

\section{Mean field approach}
\begin{figure}
\includegraphics[width=0.8\columnwidth]{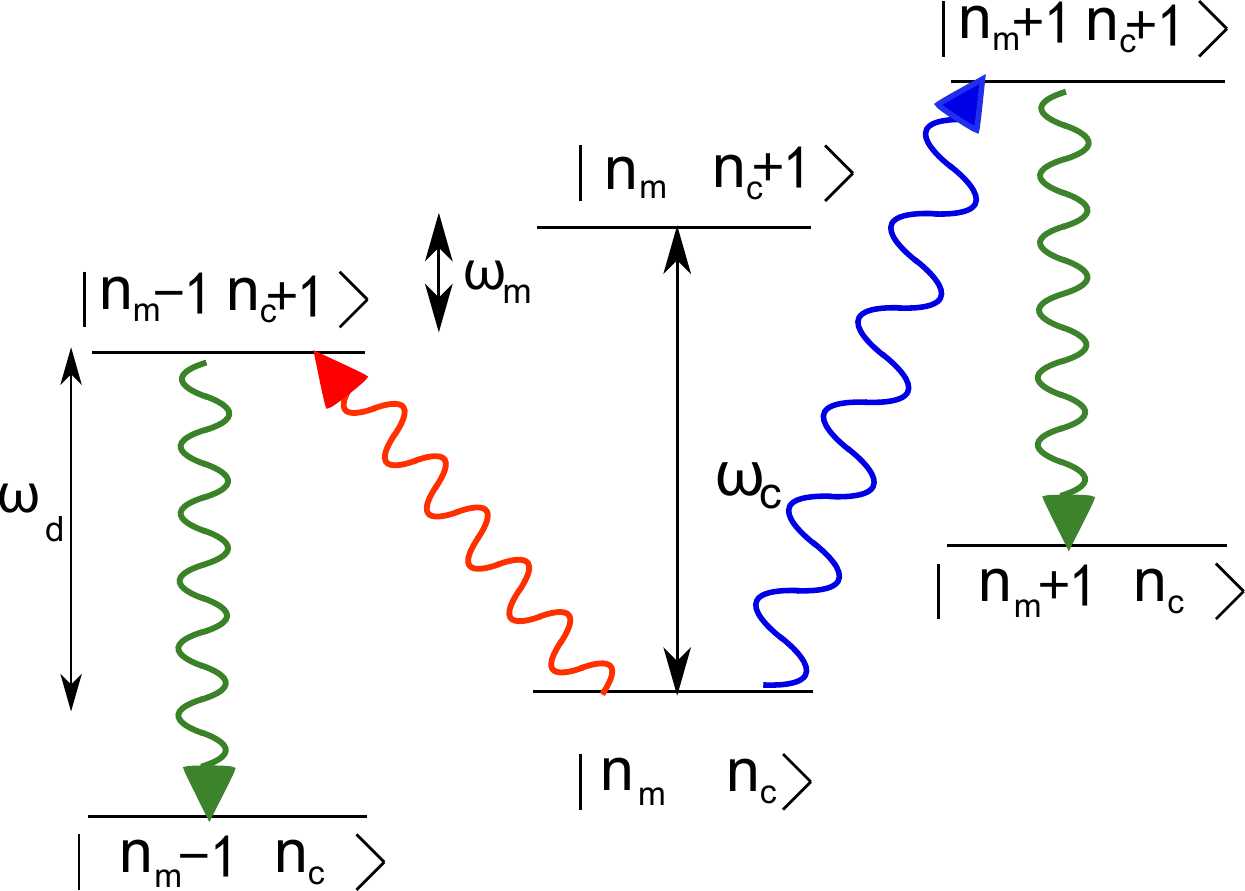}
\caption{Cooling (heating) process. The cavity is driven  with a frequency $\omega_d=\omega_c-\omega_m$ ($\omega_d=\omega_c+\omega_m$). The drive does not allow a transition from $|n_m, n_c\rangle\rightarrow |n_m, n_c+1\rangle$ but allow the transition from  $|n_m, n_c\rangle\rightarrow |n_m-1, n_c+1\rangle$ ($|n_m, n_c\rangle\rightarrow |n_m+1, n_c+1\rangle$). The cavity relaxes then to the state $|n_m-1, n_c\rangle$ ($|n_m+1, n_c\rangle$) resulting into cooling (heating) of the mechanical resonator.  }
\label{Fig:cooling}
\end{figure}

We consider  an electromagnetic cavity with frequency $\omega_c$ and linewidth $\kappa$ coupled to a mechanical resonator with frequency $\omega_m$ and linewidth $\gamma$. The number of phonons in the cavity $n_c$ is coupled to the vibration amplitude of the mechanical resonator $\hat{x}$ via a radiation-pressure type coupling $g$. In addition the number of photons $n_c$ is coupled to the number of phonons $n_m$  in the mechanical resonator through a cross-Kerr coupling  $g_{ck}$ (Fig.~\ref{Fig:schema}). The Hamiltonian of the system is ($\hbar=1$)
\begin{equation}
  H=\omega_c a^\dagger a+\omega_m b^\dagger b-g a^\dagger a (b^\dagger+b) -g_{ck} a^\dagger a b^\dagger b,
\end{equation}
where $a$ and $b$  are the annihilation operators of the cavity and the mechanical resonator, respectively. We treat the interactions with a mean-field  (MF) approach.  Within this frame, the radiation-pressure interaction becomes
\begin{multline}\label{eq:rpmf}
  g a^\dagger a(b^\dagger+b)=g\bigg [(\langle a^\dagger \rangle a+ \langle a \rangle a^\dagger-\langle a^\dagger a\rangle)(b^\dagger+b)\\
  + (a^\dagger a-\langle a\rangle a^\dagger- \langle a^\dagger \rangle a) \langle b^\dagger+b\rangle\bigg],
\end{multline}
where $\langle A\rangle$ stands for the average of $A$ over the static nonequilibrium state of the system (mean field). The negative terms  in Eq.~\eqref{eq:rpmf} are included to suppress double counting.  The first line of Eq.~\eqref{eq:rpmf} describes exchange processes between the resonator and the cavity while the second line gives a frequency shift of the cavity which is proportional to the average displacement of the resonator. This decomposition allows us to find the usual results of the weak radiation pressure coupling \cite{massel_microwave_2011,PhysRevA.31.3761}. In MF, the cross-Kerr coupling becomes
\begin{multline}
  g_{ck}a^\dagger a b^\dagger b=g_{ck}\bigg[\langle a^\dagger a \rangle b^\dagger b+
  \langle b^\dagger b \rangle a^\dagger a\\+\langle b^\dagger a  \rangle b a^\dagger+\langle b a^\dagger  \rangle b^\dagger a+\langle b a  \rangle b^\dagger a^\dagger+\langle b^\dagger a^\dagger  \rangle b a\bigg].
\end{multline}
The term  $\langle a^\dagger a \rangle b^\dagger b$ ($\langle b^\dagger b \rangle a^\dagger a$)  describes a Hartree-like  interaction between the resonator and the cavity while the other terms describe exchange processes between the  resonator and the cavity.
Thus we can rewrite the  Hamiltonian as
\begin{multline}
  H=[\omega_c-g_{ck}\langle b^\dagger b\rangle]a^\dagger a+[\omega_m-g_{ck}\langle a^\dagger a\rangle]b^\dagger b\\
  -G[\langle a^\dagger\rangle a b^\dagger+\langle a\rangle a^\dagger b^\dagger]-G^*[\langle a^\dagger \rangle ab+\langle a\rangle a^\dagger b]\\
  +g[ (a^\dagger a-\langle a\rangle a^\dagger-\langle a^\dagger \rangle a)\langle b^\dagger+b\rangle-\langle a^\dagger  a \rangle(b^\dagger +b)],
\end{multline}
where the expectation values of the different operators have to be determined self-consistently within the MF picture and $G= g+g_{ck}\langle b\rangle$. We assume the usual experimental situation where $\omega_c\gg\omega_m$ and where the cavity is driven with a coherent field of strength $f_p$ oscillating at frequency $\omega_p=\omega_c+\Delta$.  Using the input-output formalism \cite{PhysRevA.31.3761} the equations of motion are
\begin{eqnarray}
  \dot{a}&=&-i[-\Delta-g_{ck} \langle b^\dagger b \rangle]a-\frac{\kappa}{2}a+\sqrt{\kappa} f_p \nonumber \\
   \label{eq:dota} &+&iG^*\langle a\rangle b+iG\langle a\rangle b^\dagger -i g\langle b^\dagger + b\rangle [ a-\langle a\rangle]\\
  \dot{b}&=&-i[\omega_m-g_{ck}\langle a^\dagger a\rangle]b-\frac{\gamma}{2}b+\sqrt{\gamma} b_{in} \nonumber \\
    \label{eq:dotb} &&+iG\langle a^\dagger\rangle a+iG\langle a \rangle a^\dagger-i g\langle a^\dagger a \rangle,
\end{eqnarray}
Here we have written the cavity operator $a$ in a frame rotating with frequency $\omega_p$  neglecting the fast oscillatiing terms. We define $b_{in}$ to be the thermal input of the resonator satisfying $\langle b_{in}(t)\rangle=0$ and $\langle b_{in}^\dagger(t)b_{in}(t')\rangle=n^{th}\delta(t-t')$, where $n^{th}$ is the number of phonons in the thermal bath damping the resonator.  We split the cavity and the mechanical operators into a sum of  coherent  and   fluctuation parts, i.e., $a\equiv \delta a+\alpha$ and $b\equiv \delta b+\beta$ with $\alpha=\langle a \rangle$, $\beta=\langle b \rangle$ and $\langle \delta a\rangle=\langle \delta b\rangle=0 $. As usual, we assume that $\alpha$ and $\beta$ oscillate at the same frequency as the coherent drive so that $\dot{\alpha}=\dot{\beta}=0$.  With these approximations, the solutions of Eqs. (\ref{eq:dota},~\eqref{eq:dotb}) are
\begin{eqnarray}\label{eq:nc0}
 \alpha&=&\frac{\sqrt{\kappa} f_p}{\frac{\kappa}{2}-i[\Delta-g_{ck}\langle b^\dagger b \rangle -(G^* \beta+G\beta^*)]},\\\label{eq:nm0}
\beta&=&\frac{i(2G-g)|\alpha|^2-ig\langle \delta a^\dagger  \delta a\rangle}{\gamma/2+i(\omega_m-g_{ck}\langle a^\dagger a\rangle))}.
\end{eqnarray}
In the derivation of Eqs. (\ref{eq:nc0},~\ref{eq:nm0}), we have assumed, in agreement with what is usually done in the optomechanical literature (see e.g. \cite{Aspelmeyer:2013vra}),
$ \Delta + g \langle b^\dagger + b \rangle \approx\Delta$.
The equations of motion for the fluctuations in the Fourier space are given by 
\begin{eqnarray}\label{eq:da}
\left[\frac{\kappa}{2}- i(\omega+\tilde{\Delta})\right]\delta a&=&iG\alpha\delta b^\dagger +iG^*\alpha\delta b\label{eq:db}\\
\left[\frac{\gamma}{2}-i(\omega-\tilde{\omega}_m)\right] \delta b&=& iG\alpha^*\delta a+iG\alpha\delta a^\dagger+\sqrt{\gamma} b_{in},
\end{eqnarray}
where $\tilde{\Delta}=\Delta+g_{ck}\langle b ^\dagger  b\rangle$ and  $\tilde{\omega}_m=\omega_m-g_{ck}\langle a ^\dagger  a\rangle $.  The effect of the thermal drive $b_{in}$ on the response of the cavity is mediated by the coupling $G$. Through this coupling the oscillations of the mechanical resonator produce sideband peaks at $\omega_d\pm\tilde \omega_m$ in the cavity response. They allow for the exchange of energy between the  cavity and the resonator when $\Delta \approx \pm\omega_m$  \cite{PhysRevLett.97.243905,Teufel}. These processes are depicted in  Fig.~\ref{Fig:cooling}.  For $\Delta \approx -\omega_m$ the system is in the red sideband regime and one can transfer energy from the resonator to the cavity, thus the mechanical resonator is damped and cooled. For $\Delta \approx \omega_m$, the system is in the blue sideband regime and one can transfer energy from the  cavity to the resonator, thus the mechanical resonator is excited and heated. 
In order to find the correction to the damping, we solve the response function of $\delta a$  for the thermal input $\delta b_{in}$. We find that it  is a Lorentzian function peaked at $\tilde{\omega}_m+\omega_{\mathrm{shift}}$ with 
\begin{multline}
\omega_{\mathrm{shift}}=-\frac{|G|^2|\alpha|^2(\tilde{\Delta}^2-\tilde{\omega}_m^2+\frac{\kappa^2}{4})}{\tilde{\omega}_m}\\\left(\frac{1}{\frac{\kappa^2}{4}+(\tilde{\omega}_m+\tilde{\Delta})^2}-\frac{1}{\frac{\kappa^2}{4}+(\tilde{\omega}_m-\tilde{\Delta})^2}\right),
 \end{multline}
and whose  linewidth is $\gamma+\Gamma_{\mathrm{opt}}$ with
\begin{multline}\label{eq:gopt}
\Gamma_{\mathrm{opt}}=|G|^2|\alpha|^2 \kappa\\\left(\frac{1}{\frac{\kappa^2}{4}+(\tilde{\omega}_m+\tilde{\Delta})^2}-\frac{1}{\frac{\kappa^2}{4}+(\tilde{\omega}_m-\tilde{\Delta})^2}\right).
\end{multline}
Integrating the Lorentzian function obtained above, we obtain the number of phonons and photons coming from the thermal vibrations of the resonator. We get \cite{PhysRevLett.99.093902}
\begin{eqnarray}\label{eq:nmth}
\langle \delta b^\dagger\delta b\rangle&=&\frac{\gamma n^{th}+\Gamma_{opt}n_{m_0}}{\gamma+\Gamma_{\mathrm{opt}}},\\\label{eq:ncth}
\langle \delta a^\dagger \delta a\rangle&=&G^2|\alpha|^2\langle \delta b^\dagger \delta b\rangle\\ \nonumber&& \left(\frac{1}{\frac{\kappa^2}{4}+(\tilde{\omega}_m+\tilde{\Delta})^2}+\frac{1}{\frac{\kappa^2}{4}+(\tilde{\omega}_m-\tilde{\Delta})^2}\right)
\end{eqnarray}
with 
\begin{eqnarray}\
n_{m_0}&=&-\frac{(\tilde{\omega}_m+\tilde{\Delta})^2+\frac{\kappa ^2}{4}}{4\tilde{\Delta}\tilde{\omega}_m}.
\end{eqnarray}
Eqs. \eqref{eq:nc0}, \eqref{eq:nm0}, \eqref{eq:nmth} and \eqref{eq:ncth} form a set of self-consistency equations and are solved in the next sections in order to find the number of phonons in the resonator and photons in the cavity. We now focus on the sidebands.

\section{optimal cooling/heating}
\begin{figure}
\includegraphics[width=1\columnwidth]{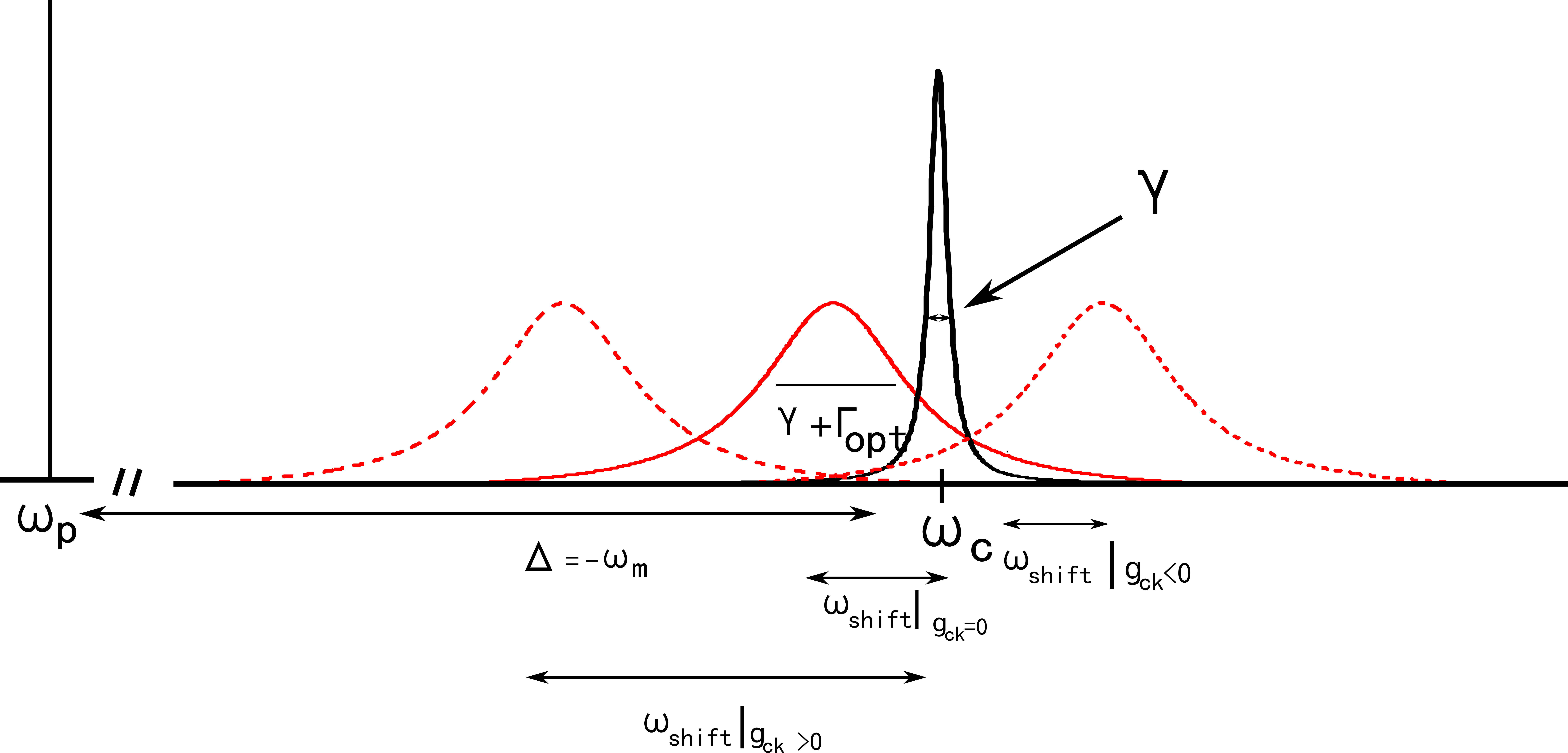}
\caption{Schematic picture of the red sideband with and without cross-Kerr coupling for $\Delta<0$. For $g_{ck}>0$ the sideband peak is shifted to lower values while for $g_{ck}<0$ the sideband peak is shifted to higher values.}
\label{Fig:sbcomp}
\end{figure}

 In order to minimize/maximize the optical damping $\Gamma_{opt}$, we set $\tilde{\Delta}=\mp\tilde{\omega}_m$ . The upper sign refers to the red sideband ($\Gamma_{opt} >0$) and the lower sign to the blue sideband ($\Gamma_{opt}<0$). In the  resolved sideband limit,   $\omega_m\gg\kappa\gg\gamma$,  the  frequency shift and the optical damping become
\begin{eqnarray} \label{eq:omegaoptrsb}
\omega_{\mathrm{shift}}&=&\mp \frac{|G|^2|\alpha|^2}{\tilde{\omega}_m}=\mp\frac{|G|^2|\alpha|^2}{\omega_m-g_{ck}\langle  a^\dagger a\rangle},\\\label{eq:gammaoptrsb}
\Gamma_{\mathrm{opt}}&=&\pm\frac{4|G|^2|\alpha|^2}{\kappa}.
\end{eqnarray}
The result for the optical damping Eq. \eqref{eq:gammaoptrsb}, is identical to  the one usually obtained    in optomechanics in the absence of the cross-Kerr coupling. The effect of $g_{ck}$ shows up only in the frequency shift which now depends on the number of coherent and thermal photons in the cavity Eq.~\eqref{eq:omegaoptrsb}. In Fig.~\ref{Fig:sbcomp} we show a schematic picture of what happens to the sideband in the presence of the cross-Kerr coupling.

\begin{figure}
\includegraphics[width=1\columnwidth]{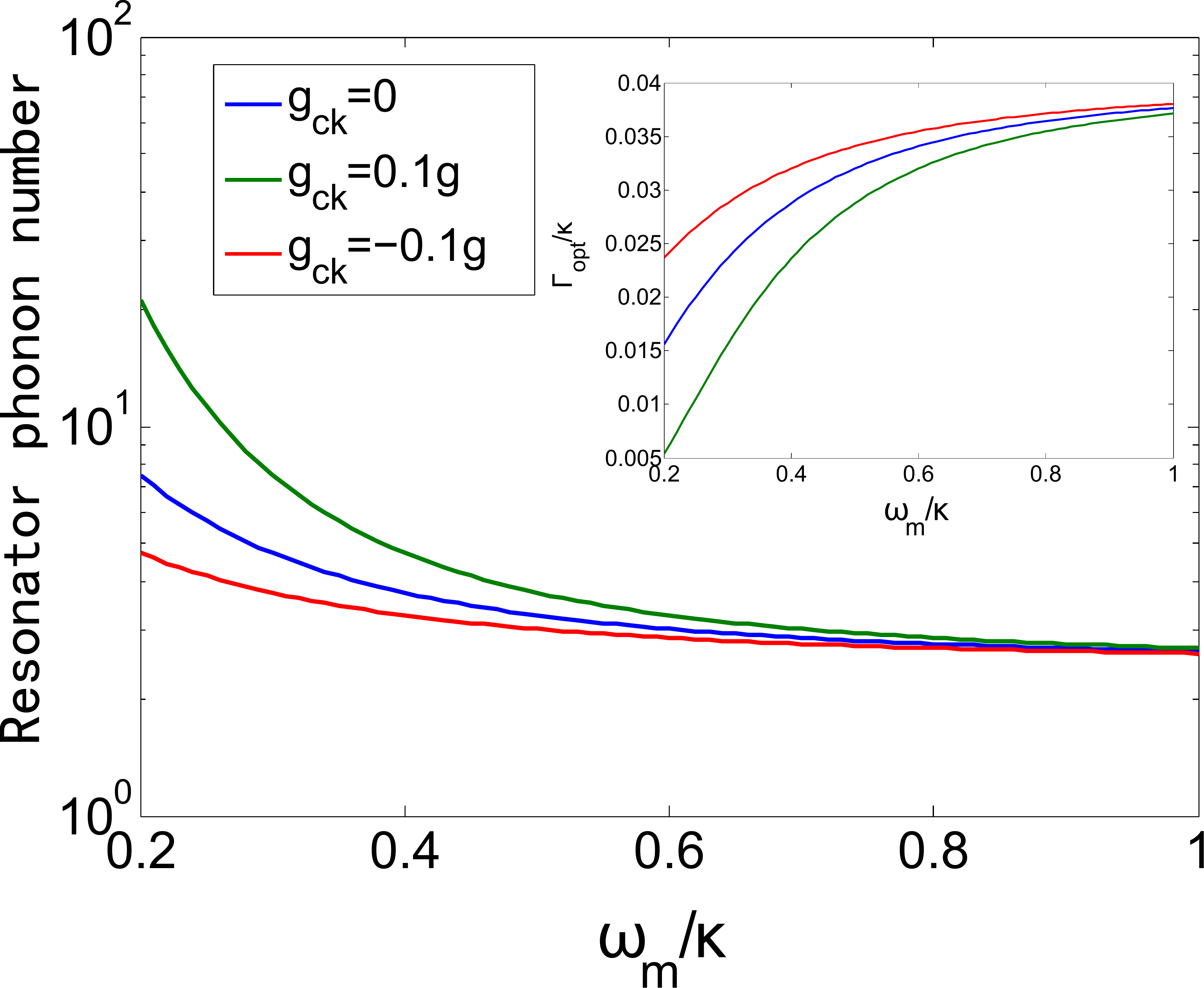}
\caption{Steady-state phonon number in the resonator and $\Gamma_{\mathrm{opt}}$ (inset) as a function of the ratio $\omega_m/\kappa$ at the red sideband for the optimal case  $\tilde{\Delta}=-\tilde{\omega}_m$ with $\gamma=10^{-3}\kappa$, $g=10^{-2}\kappa$. The number of photons pumped into the cavity is fixed to $|\alpha|^2=100$ and the bath temperature corresponds to $n^{th}=100$. }
\label{Fig:optimal}
\end{figure}

\begin{figure}
\includegraphics[width=1\columnwidth]{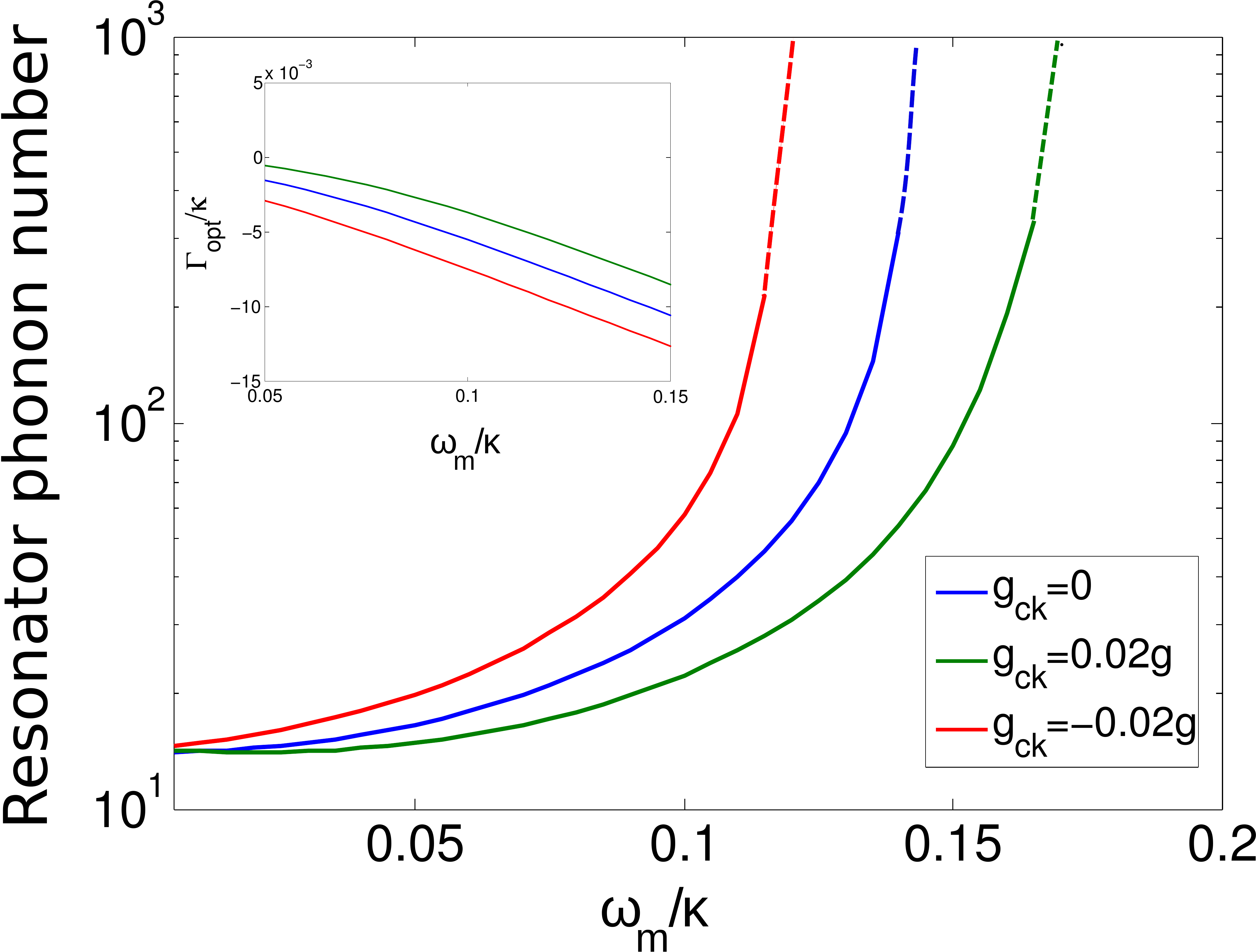}
\caption{Steady-state phonon number in the resonator and $\Gamma_{\mathrm{opt}}$ (inset) as a function of the ratio $\omega_m/\kappa$ at the blue sideband for the optimal case $\tilde{\Delta}=\tilde{\omega}_m$ with $\gamma=10^{-2}\kappa$, $g=10^{-2}\kappa$. The number of photons pumped into the cavity is fixed to $|\alpha|^2=100$ and the bath temperature corresponds to $n^{th}=10$. The dashed lines indicate the onset of the parametric instability for $\Gamma_{\rm opt} = -\gamma$.}
\label{Fig:optimalblue}
\end{figure}

In the Doppler limit  ($\omega_m\leq \kappa$) the frequency shift  and optical damping are given by 
\begin{eqnarray}\label{eq:omeganrsd}
\omega_{\mathrm{shift}}&=&\mp4|G|^2|\alpha|^2\frac{\omega_m-g_{ck}\langle  a^\dagger a\rangle}{\frac{\kappa^2}{4}+4(\omega_m-g_{ck}\langle  a^\dagger a\rangle)^2}\\ \label{eq:gammanrsd}
\Gamma_{\mathrm{opt}}&=&\pm\frac{4|G|^2|\alpha|^2}{\kappa}\frac{4({\omega}_m-g_{ck}\langle  a^\dagger a\rangle)^2}{\frac{\kappa^2}{4}+4({\omega}_m-g_{ck}\langle  a^\dagger a\rangle)^2}.
\end{eqnarray}
Now both the frequency shift and the optical damping   depend  on the cross-Kerr coupling. In Figs.~\ref{Fig:optimal}-\ref{Fig:optimalblue} we plot the number of phonons and the optical damping as a function of  $\omega_m/\kappa$ for the red sideband in the Doppler limit. Since the cross-Kerr coupling shifts the mechanical frequency, the value of $\Gamma_{\rm opt}$ is shifted as well. The sign of the shift is given by the sign of $g_{ck}$.  Otherwise we recover the cooling of the resonator for the red sideband  (Fig.~\ref{Fig:optimal}) and the parametric instability when $\Gamma_{\rm opt}=-\gamma$  for the blue sideband  (Fig.~\ref{Fig:optimalblue}).

\section{ Case with $\Delta=\omega_m.$} In experiments the parameter one can tune directly is the detuning $\Delta$ and not $\tilde{\Delta}$ as it can be difficult to set $\tilde{\Delta}=\mp\tilde{\omega}_m$ for each value of $|\alpha|$ as the pump strength is varied. Therefore, another regime we consider is the case where $\Delta=\mp\omega_m$, i.e, setting $\tilde{\Delta}=\mp\omega_m+g_{ck}\langle b^\dagger  b \rangle$. In this case the frequency shift and optical damping in the red (upper sign) and in the blue (lower sign) sideband  become
\begin{eqnarray}
\omega_{\mathrm{shift}}&=&\mp\frac{|G|^2|\alpha|^2}{\tilde{\omega}_m}\nonumber\\
&&\frac{[\kappa^2/4-2g_{ck}\omega_m(\langle b^\dagger b \rangle -\langle a^\dagger a \rangle)]}{[g_{ck}^2(\langle b^\dagger b \rangle -\langle a^\dagger a \rangle)^2+\kappa^2/4]},\\
\Gamma_{\mathrm{opt}}&=&\pm\frac{|G|^2|\alpha|^2\kappa}{g_{ck}^2(\langle b^\dagger b \rangle -\langle a^\dagger a \rangle)^2+\kappa^2/4}.
\end{eqnarray}

\begin{figure}
\includegraphics[width=0.9\columnwidth]{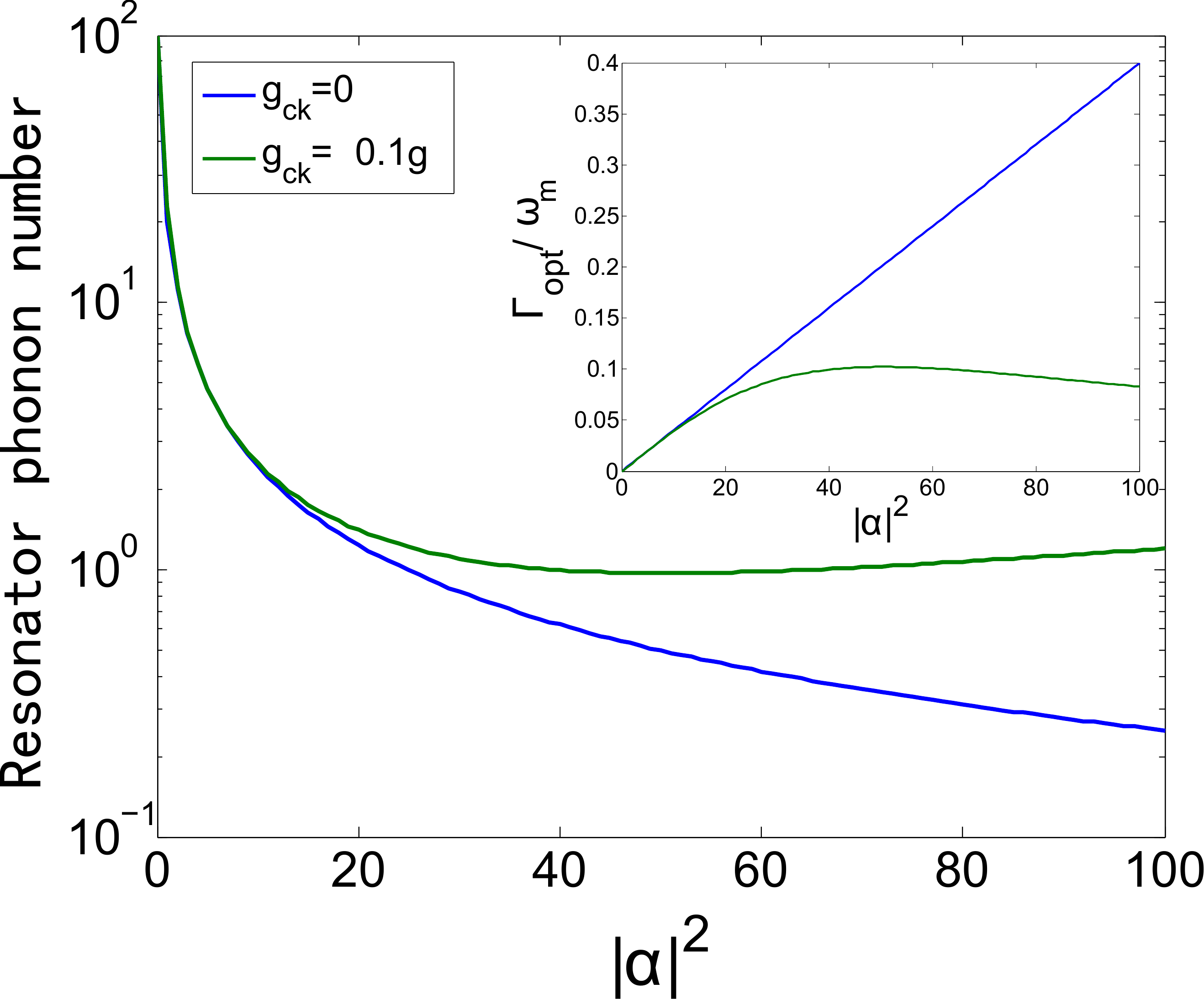}
\caption{Steady-state phonon number in the resonator and the optical damping $\Gamma_{\mathrm{opt}}$   (inset) as a function of  the number of photons pumped into the cavity in the case where $\Delta=-\omega_m$. The values for the parameters are $\gamma=10^{-3}\omega_m$,  $\kappa=10^{-1}\omega_m$, $g=10^{-2}\omega_m$ and the bath temperature corresponds to $n^{th}=100$.}
\label{Fig:gammaeff}
\end{figure}

\begin{figure}
\includegraphics[width=0.9\columnwidth]{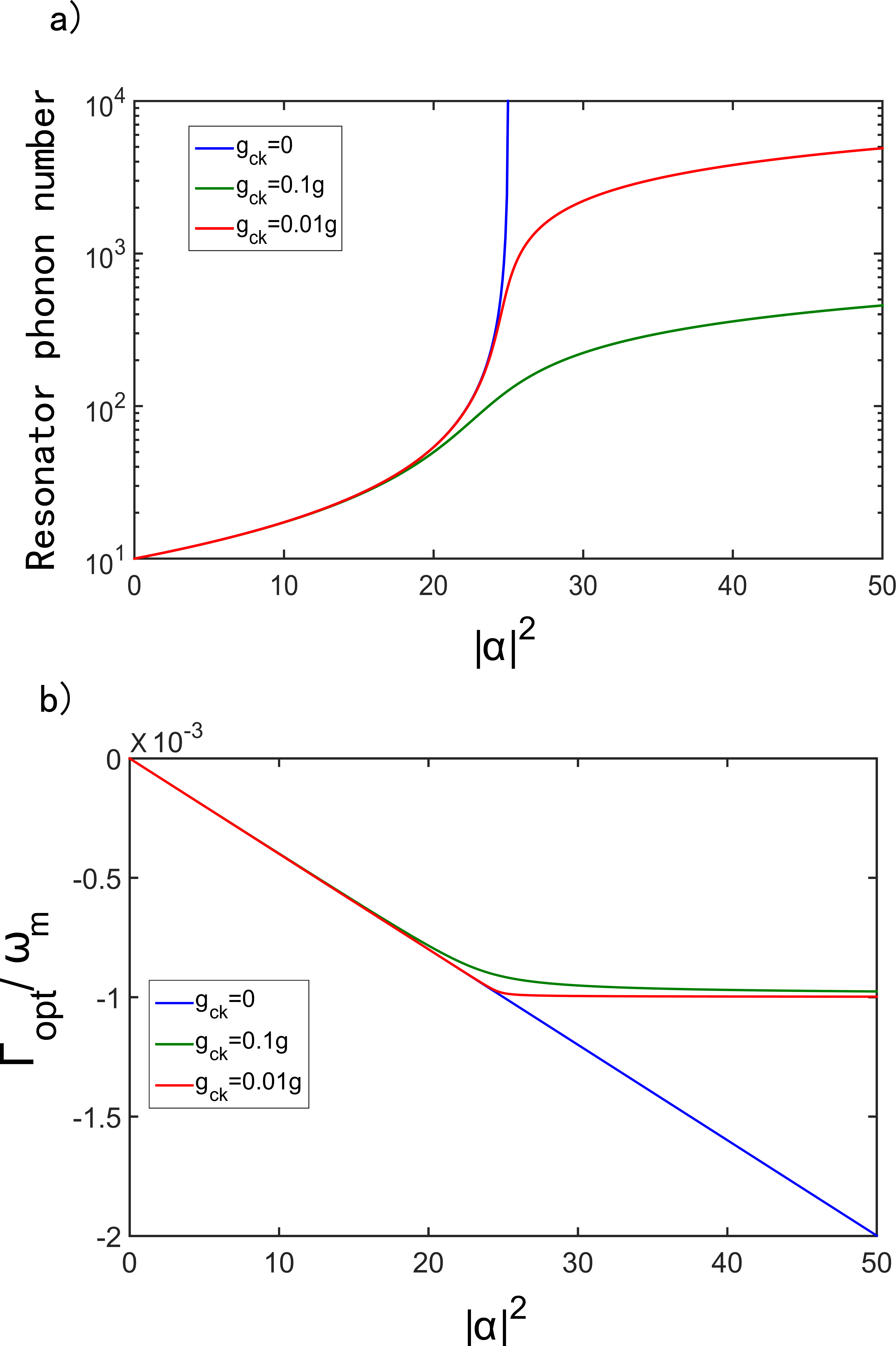}
\caption{a) Steady-state phonon number in the resonator and b) the optical damping $\Gamma_{\mathrm{opt}}$   as a function of  the number of photons pumped into the cavity in the case where $\Delta=\omega_m$. The values for the parameters are $\gamma=10^{-3}\omega_m$,  $\kappa=10^{-1}\omega_m$, $g=10^{-2}\omega_m$ and the bath temperature corresponds to $n^{th}=10$.}
\label{Fig:gammaeffblue}
\end{figure}

In Figs.~\ref{Fig:gammaeff} and  \ref{Fig:gammaeffblue} the steady-state phonon number and the optical damping are plotted as a function of  the number of photons pumped into the cavity, for the red and blue sidebands respectively. For the red sideband (Fig.~\ref{Fig:gammaeff}) the optical damping increases with increasing $\alpha$. When $g_{ck}^2|\alpha|^2\gg  \kappa^2/4$ the optical damping becomes inversely proportional to the number of photons pumped into the cavity, consequently, the cooling deteriorates when pumping more phonons in the cavity. 

In the blue sideband (Fig.~\ref{Fig:gammaeffblue}) the main effect of a small cross-Kerr coupling is to limit the instability to a finite number of phonons, $\langle b^\dagger b \rangle \approx \sqrt{\kappa/\gamma} |G| |\alpha|/|g_{ck}|+|\alpha|^2$. This thus competes with the usual limitation coming from the intrinsic (Duffing) nonlinearity of the resonator.

\begin{figure}
\includegraphics[width=0.9\columnwidth]{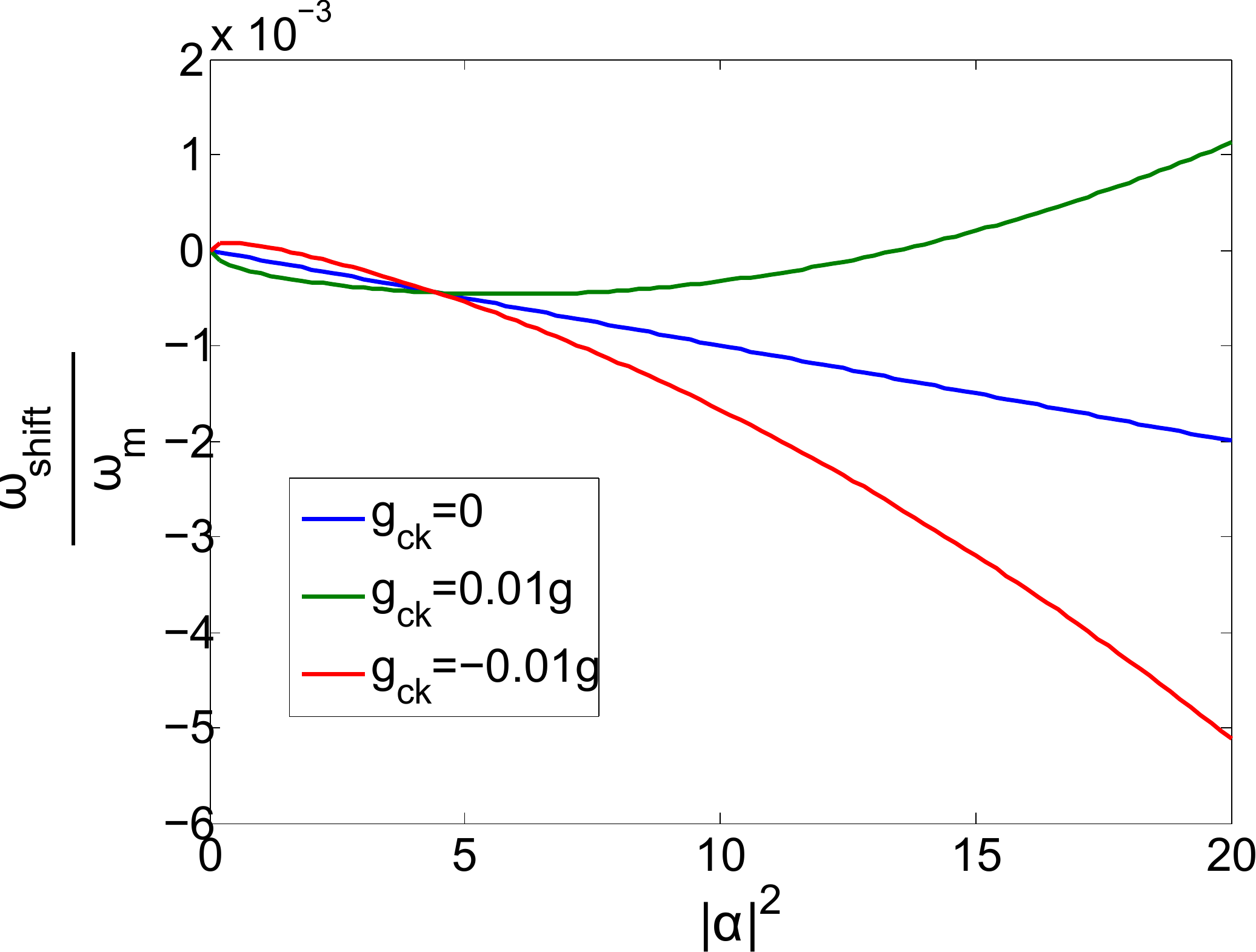}
\caption{Frequency shift as a function of  the number of photons pumped into the cavity when $\Delta=-\omega_m$ with $\gamma=10^{-3}\omega_m$ $\kappa=10^{-1}\omega_m$, $g=10^{-2}\omega_m$ and the bath temperature corresponds to $n^{th}=100$.}
\label{Fig:omegaeff}
\end{figure}

\begin{figure}
\includegraphics[width=0.9\columnwidth]{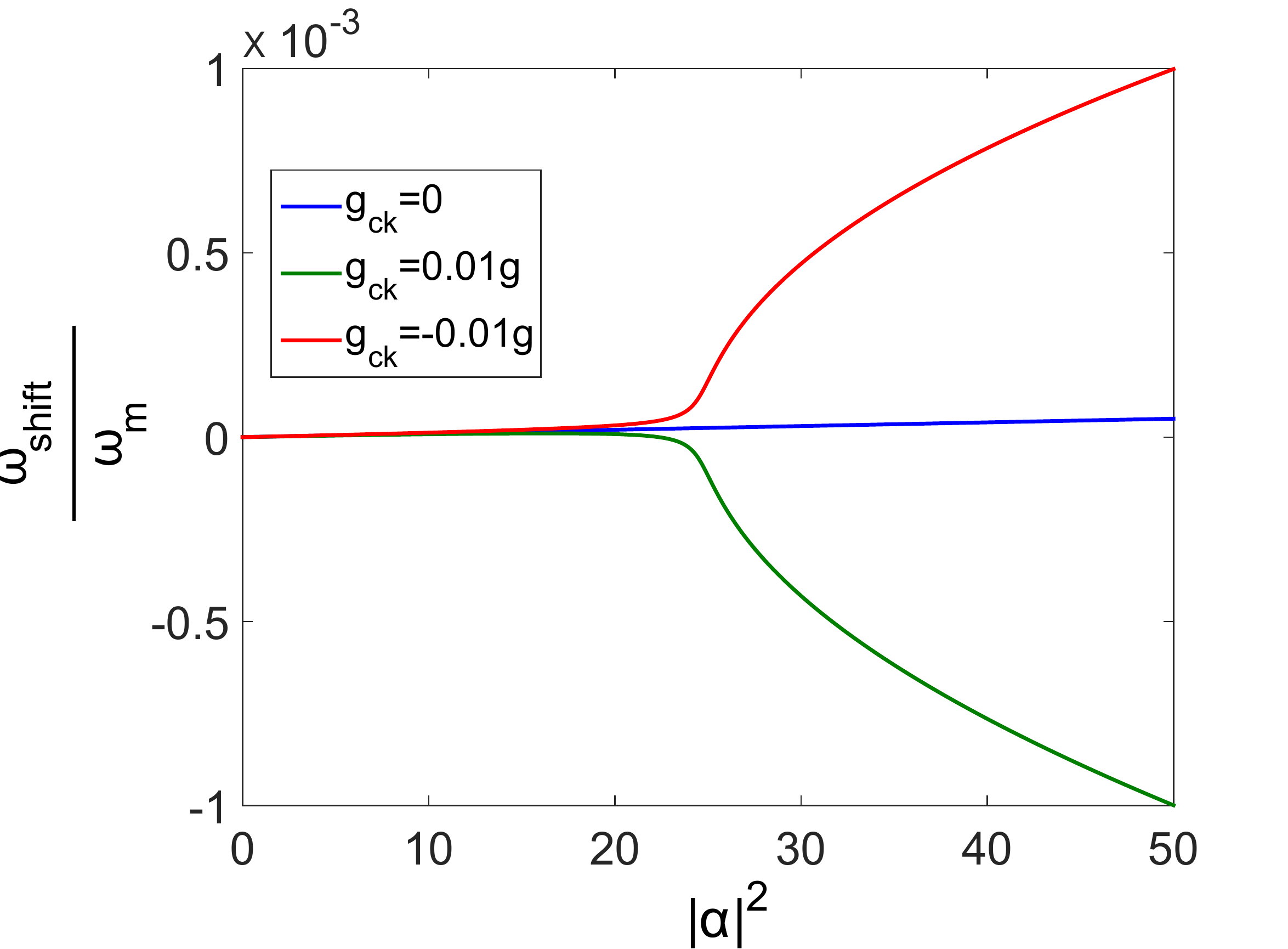}
\caption{Frequency shift as a function of  the number of photons pumped into the cavity when $\Delta=\omega_m$ with $\gamma=10^{-3}\omega_m$ $\kappa=10^{-1}\omega_m$, $g=10^{-2}\omega_m$ and the bath temperature corresponds to $n^{th}=100$. }
\label{Fig:omegaeffblue}
\end{figure}

In Figs.~\ref{Fig:omegaeff}-\ref{Fig:omegaeffblue} we plot the frequency shift  as a function of the number of photons pumped into the cavity for the red and blue sidebands. For the red sideband when $g_{ck}>0$ ($g_{ck}<0$) the frequency shift increases (decreases) as $\alpha$ increases  until  $\alpha\approx \langle b^\dagger b\rangle$ after which it increases (decreases). For the blue sideband when $g_{ck}>0$ the frequency shift decreases while for for $g_{ck}<0$ it increases. The difference at  small $\alpha$ between the red and blue sidebands arises from the fact that in the red sideband when pumping more photons into the cavity the cooling improves, thus the number of  phonons in the mechanical resonator decreases, making it possible to have a number of photons in the cavity of the same order and larger than  the number of phonons in the resonator. 


\section{Conclusion}
In conclusion, we have solved  the dynamics of a mechanical resonator coupled to an electromagnetic cavity via a radiation  pressure coupling and  a cross-Kerr coupling using a mean field approach. We have shown that the cross-Kerr coupling shifts the frequency of the mechanical resonator and of the optical cavity, the shift depending on the number of photons in the cavity and phonons in the resonator. In addition, we have shown that when the detuning of the pump is equal to the frequency of the mechanical resonator the variation of the optical damping saturates instead of being linearly dependent on the number of phonons pumped into the cavity.

 This work was supported by the European Research Council (Grant No. 240362-Heattronics) and the Academy of Finland.


\end{document}